\documentclass[12pt]{article}%
\usepackage{citesort}
\usepackage{amsmath}
\usepackage{graphicx}%
\usepackage{amsfonts}%
\usepackage{amssymb}

\begin{document}

\title{Long-range multiplicity correlations in \\proton-proton collisions}
\author{Adam Bzdak\thanks{e-mail: Adam.Bzdak@ifj.edu.pl}\\Institute of Nuclear Physics, Polish Academy of Sciences\\Radzikowskiego 152, 31-342 Krakow, Poland}
\maketitle

\begin{abstract}
The forward-backward long-range multiplicity correlations in proton-proton
collisions are investigated in the model with two independent sources of
particles: one left- and one right-moving wounded nucleon. A good agreement
with the UA5 collaboration proton-antiproton data at the c.m. energy of $200$
GeV is observed. For comparison the model with only one source of particles is
also discussed.

\vskip 0.6cm


\end{abstract}

\section{Introduction}

\begin{figure}[h]
\begin{center}
\includegraphics[scale=1.5]{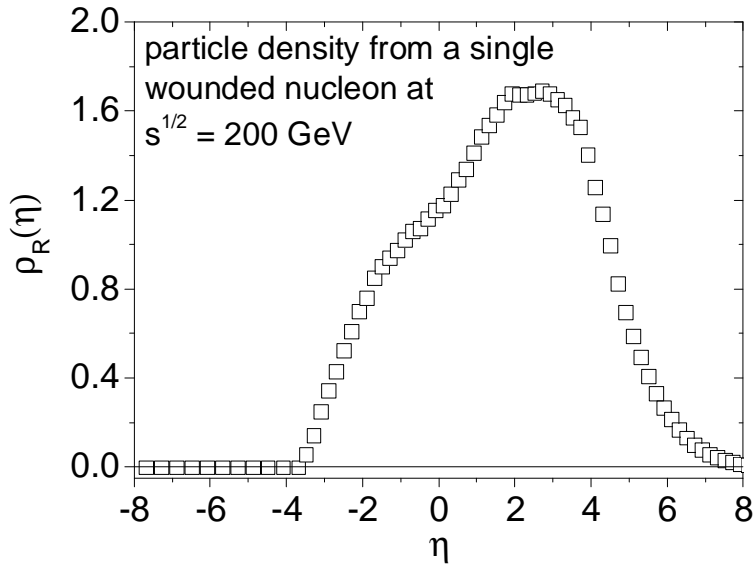}
\end{center}
\caption{The pseudorapidity density of produced charged particles from a
right-moving wounded nucleon at $\sqrt{s}=200$ GeV c.m. energy. $\rho_{R}%
(\eta)$ substantially feeds into the opposite hemisphere which implies
specific long-range forward-backward multiplicity correlations.}%
\label{fig_wn}%
\end{figure}

Recently, the pseudorapidity particle density from a wounded
nucleon\footnote{The wounded nucleon is the one which underwent at least one
inelastic collision \cite{WNM}.} was determined by analysing the PHOBOS data
\cite{PHO-dAu} on deuteron-gold collisions at $\sqrt{s}=200$ GeV in the
framework of the wounded nucleon \cite{ff-bc} and the wounded quark-diquark
\cite{ff-bb} models. The obtained fragmentation function\footnote{In this
picture all \textit{soft} particles are produced independently from left- and
right-moving wounded nucleons. It is very similar to the assumption of
independent hadronization of strings in the dual parton model \cite{DPM}.} has
two characteristic features. It is peaked in the forward direction and it
substantially feeds into the opposite hemisphere, as shown in Fig.
\ref{fig_wn}. In Ref. \cite{ff-Ryb} very similar shape of the contribution
from a wounded nucleon was found at SPS energy of $\sqrt{s}=17.3$ GeV. The
possible explanation of the main features of the wounded nucleon fragmentation
function was proposed in Ref. \cite{bibzpe} in the model based on the
bremsstrahlung mechanism \cite{Stod}.

\begin{figure}[h]
\begin{center}
\includegraphics[scale=1.1]{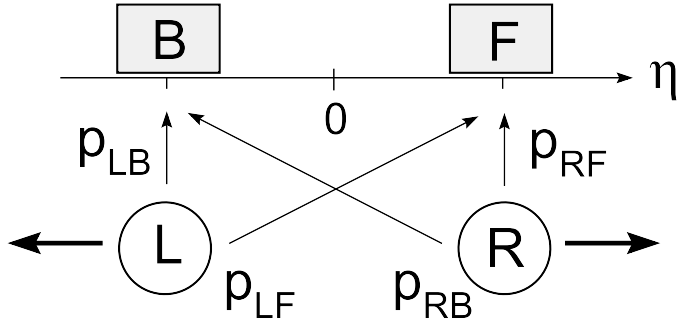}
\end{center}
\caption{The model of soft particle production in proton-proton collisions
with two independent sources of particles: left- and right-moving wounded
nucleons. The arrows indicate that each wounded nucleon may populate particles
into both pseudorapidity intervals with appropriate probabilities.}%
\label{fig_two}%
\end{figure}

It is interesting to notice that the picture in which the wounded nucleon
populates particles into the opposite hemisphere implies specific long-range
forward-backward multiplicity correlations. This problem will be investigated
here in the context of the UA5 $p\bar{p}$ forward-backward multiplicity
correlation data at $\sqrt{s}=200$ GeV \cite{UA5-fb}. Namely, we will test the
model with two independent sources of particles, see Fig. \ref{fig_two}, with
the wounded nucleon fragmentation function shown in Fig. \ref{fig_wn}. For
comparison, we will also study the model in which particles are produced from
only one source of particles e.g., a single string spanned between two wounded
nucleons. 

Our main conclusion is that the model with two independent sources of
particles and the wounded nucleon fragmentation function extracted from the
PHOBOS $d-Au$ data is fully consistent with the UA5 $p\bar{p}$
forward-backward multiplicity correlation data at $\sqrt{s}=200$ GeV. At the
same time we conclude that the model with only one source of particles is in a
very clear disagreement with the data.

Let us emphasize here that the idea of the multicomponent model of soft
particle production is not new and was successfully applied to the
forward-backward multiplicity correlations data by many authors
\cite{DPM,FK,DDD,BBP,Giov,Braun,Brogueira}. Our model with two sources of
particles has very much in common with the dual parton model \cite{DPM} or the
two-chain dual model with non-zero asymmetry of each chain \cite{FK,DDD}. For
other approaches see Refs. \cite{Chou,Lim}.

In the next section basic formulae are introduced. In section $3$ we present
our model in detail and derive for $pp$ collisions the analytical expressions
for the correlation coefficient and the functional relation between the
average number of particles in the backward interval at a given number of
particles in the forward one. We also discuss the limit of one-source model.
In section $4$ our results are tested using UA5 $p\bar{p}$ forward-backward
multiplicity correlation data. Our conclusions are listed in the last section,
where also some comments are included.

\section{General formulae}

It is convenient to construct the generating function%
\begin{equation}
H\left(  z_{B},z_{F}\right)  =\sum\nolimits_{n_{B},n_{F}}P\left(  n_{B}%
,n_{F}\right)  z_{B}^{n_{B}}z_{F}^{n_{F}}, \label{H1}%
\end{equation}
where $P\left(  n_{B},n_{F}\right)  $ is the probability in $pp$ collisions to
find $n_{B}$ particles in $B$ interval and $n_{F}$ particles in $F$ interval,
see Fig. \ref{fig_two}. It is worth to notice that the generating function
(\ref{H1}) contains all information about the multiplicities in $B$ and $F$.

The correlation coefficient (or correlation strength) is defined as%
\begin{equation}
b=\frac{\left\langle n_{B}n_{F}\right\rangle -\left\langle n_{B}\right\rangle
\left\langle n_{F}\right\rangle }{\left\langle n_{F}^{2}\right\rangle
-\left\langle n_{F}\right\rangle ^{2}}, \label{b_def}%
\end{equation}
where $n_{B}$ and $n_{F}$ are event by event particle multiplicities in $B$
and $F$ intervals, respectively. If the number of particles in $B$ interval
does not dependent on the number of particles in $F$ i.e., $\left\langle
n_{B}n_{F}\right\rangle =\left\langle n_{B}\right\rangle \left\langle
n_{F}\right\rangle $ we have $b=0$. On the other hand, if $n_{B}=n_{F}$ in
every event then $b=1$ (maximum correlation). Using definition (\ref{H1}) the
correlation coefficient $b$ can be expressed by the appropriate derivatives of
the generating function%
\begin{align}
\left\langle n_{B}n_{F}\right\rangle -\left\langle n_{B}\right\rangle
\left\langle n_{F}\right\rangle  &  =\left[  \frac{\partial^{2}H}{\partial
z_{B}\partial z_{F}}-\frac{\partial H}{\partial z_{B}}\frac{\partial
H}{\partial z_{F}}\right]  _{z_{B}=1,z_{F}=1},\nonumber\\
\left\langle n_{F}^{2}\right\rangle -\left\langle n_{F}\right\rangle ^{2}  &
=\left[  \frac{\partial^{2}H}{\partial z_{F}^{2}}+\frac{\partial H}{\partial
z_{F}}-\left(  \frac{\partial H}{\partial z_{F}}\right)  ^{2}\right]
_{z_{B}=1,z_{F}=1}. \label{b_poch}%
\end{align}

It is also interesting to study the functional relation between the average
number of particles $\left\langle n_{B}\right\rangle $ in $B$ interval under
the condition of $n_{F}$ particles in $F$ interval%
\begin{equation}
\left\langle n_{B}\right\rangle |_{n_{F}}=\frac{\sum\nolimits_{n_{B}}%
n_{B}P\left(  n_{B},n_{F}\right)  }{\sum\nolimits_{n_{B}}P\left(  n_{B}%
,n_{F}\right)  }, \label{line_def}%
\end{equation}
where the numerator and denominator can be expressed by the derivatives of the
generating function (\ref{H1})
\begin{align}
\sum\nolimits_{n_{B}}P\left(  n_{B},n_{F}\right)   &  =\left.  \frac{1}%
{n_{F}!}\frac{\partial^{n_{F}}H\left(  z_{B},z_{F}\right)  }{\partial
z_{F}^{n_{F}}}\right|  _{z_{B}=1,z_{F}=0},\nonumber\\
\sum\nolimits_{n_{B}}n_{B}P\left(  n_{B},n_{F}\right)   &  =\left.
\frac{1}{n_{F}!}\frac{\partial}{\partial z_{B}}\frac{\partial^{n_{F}}H\left(
z_{B},z_{F}\right)  }{\partial z_{F}^{n_{F}}}\right|  _{z_{B}=1,z_{F}=0}.
\label{line_poch}%
\end{align}

In the next section we calculate (\ref{b_def}) and (\ref{line_def}) in two
models of particle production.

\section{Model}

The schematic view of our model is presented in Fig. \ref{fig_two}. We assume
that in $pp$ collisions all soft particles are produced from two independent
wounded nucleons\footnote{The detailed discussion of this assumption and its
successful applications can be found in Refs.
\cite{ff-bc,ff-bb,ff-Ryb,fb-star}.}, which populate particles according to the
fragmentation function\footnote{In the c.m. frame a contribution from the
left-moving wounded nucleon $\rho_{L}(\eta)=\rho_{R}(-\eta)$.} presented in
Fig. \ref{fig_wn}. Additionally, we assume that in $pp$ collisions the
multiplicity distribution in the combined interval $B+F$ is described be the
negative binomial (NB) distribution%
\begin{equation}
P_{NB}(n,\bar{n},k)=\frac{\Gamma(n+k)}{\Gamma(n+1)\Gamma(k)}\left(
\frac{\bar{n}}{k}\right)  ^{n}\left(  1+\frac{\bar{n}}{k}\right)  ^{-n-k},
\label{nbd}%
\end{equation}
where $\bar{n}$ is the average multiplicity in $B+F$ and $1/k$ measures
deviation from Poisson distribution. It is obvious that $\bar{n}$ can be
calculated as
\begin{equation}
\bar{n}=\int_{B+F}\left[  \rho_{R}\left(  \eta\right)  +\rho_{L}\left(
\eta\right)  \right]  d\eta=2\int_{B+F}\rho_{R}\left(  \eta\right)  d\eta,
\label{mn}%
\end{equation}
where $\rho_{R}\left(  \eta\right)  $ and $\rho_{L}\left(  \eta\right)
=\rho_{R}\left(  -\eta\right)  $ are the pseudorapidity densities of produced
particles from the right- and left-moving wounded nucleons, respectively.

Recently, we have shown \cite{fb-star} that the generating function (\ref{H1})
in the framework of the above-mentioned model may be written as%
\begin{align}
H\left(  z_{B},z_{F}\right)   &  =\left\{  1+\frac{\bar{n}}{k}\left[
p_{LB}\left(  1-z_{B}\right)  +p_{LF}\left(  1-z_{F}\right)  \right]
\right\}  ^{-k/2}\times\nonumber\\
&  \times\left\{  1+\frac{\bar{n}}{k}\left[  p_{RB}\left(  1-z_{B}\right)
+p_{RF}\left(  1-z_{F}\right)  \right]  \right\}  ^{-k/2}, \label{H2}%
\end{align}
where $p_{RF}$ is the probability that a particle originating from the
right-moving wounded nucleon goes to $F$ interval rather than to $B$ (and
analogous for $p_{RB},p_{LB}$ and $p_{LF}$), see Fig. \ref{fig_two}. These
probabilities satisfy the following conditions
\begin{equation}
p_{LB}+p_{LF}=1,\quad p_{RB}+p_{RF}=1. \label{p_suma}%
\end{equation}
These numbers can be easily calculated using the wounded nucleon fragmentation
function. For instance, $p_{RF}$ has the form%
\begin{equation}
p_{RF}=\frac{\int_{F}\rho_{R}\left(  \eta\right)  d\eta}{\int_{B+F}\rho
_{R}\left(  \eta\right)  d\eta}. \label{p_def}%
\end{equation}

Taking (\ref{b_def}), (\ref{b_poch}) and (\ref{H2}) into account and
performing elementary calculations, the following expression for the
correlation coefficient in the model with two independent sources of particles
is obtained%
\begin{equation}
b=\frac{\bar{n}(p_{LB}p_{LF}+p_{RB}p_{RF})}{\bar{n}(p_{LF}^{2}+p_{RF}%
^{2})+k(p_{LF}+p_{RF})}. \label{b2}%
\end{equation}
Assuming that intervals $B$ and $F$ are separated enough so that $F$ can be
populated only by the right-moving nucleon and $B$ only by the left-moving one
i.e., $p_{LB}=p_{RF}=1$ and $p_{LF}=p_{RB}=0$ we obtain $b=0$. Thus, we
immediately predict the noticeable suppression of the correlation coefficient
$b$ with increasing distance between $B$ and $F$ intervals.

In the model with two independent sources of particles the relation between
the average number of particles $\left\langle n_{B}\right\rangle $ in the
backward interval $B$ at a given number of particles $n_{F}$ in the forward
interval $F$ has the form [see (\ref{line_def}), (\ref{line_poch}) and
(\ref{H2})]%
\begin{align}
\left\langle n_{B}\right\rangle |_{n_{F}}  &  =\frac{k\bar{n}p_{LB}}{2\left(
k+\bar{n}p_{LF}\right)  }\frac{_{2}F_{1}(1+k/2,-n_{F},1-n_{F}-k/2,\xi)}%
{_{2}F_{1}(k/2,-n_{F},1-n_{F}-k/2,\xi)}+\nonumber\\
&  +\frac{\bar{n}p_{RB}\left(  n_{F}+k/2\right)  }{k+\bar{n}p_{RF}}%
\frac{_{2}F_{1}(k/2,-n_{F},-n_{F}-k/2,\xi)}{_{2}F_{1}(k/2,-n_{F}%
,1-n_{F}-k/2,\xi)}, \label{nbm2}%
\end{align}
where%
\begin{equation}
\xi=\frac{p_{LF}(k+\bar{n}p_{RF})}{p_{RF}(k+\bar{n}p_{LF})},
\end{equation}
and the hypergeometric function $_{2}F_{1}(a,-n_{F},1-c,\xi)$ is defined as
\begin{equation}
_{2}F_{1}(a,-n_{F},1-c,\xi)=\frac{\Gamma(1+n_{F})}{\Gamma(a)\Gamma(c)}%
\sum\limits_{M=0}^{n_{F}}\frac{\Gamma(a+M)\Gamma(c-M)}{M!(n_{F}-M)!}\xi^{M}.
\end{equation}

For comparison we also derive the appropriate formulae in the model with only
one source of particles e.g., a single string spanned between two wounded
nucleons. These expressions can be easily obtained from (\ref{b2}) and
(\ref{nbm2}) by \textit{deactivating} one of the sources e.g., the left one.
In this case $p_{LB}=p_{LF}=0$ (thus $\xi=0$) and $p_{RB}\equiv p_{B}$ and
$p_{RF}\equiv p_{F}$ where $p_{B}+p_{F}=1$. Finally
\begin{equation}
b=\frac{p_{B}\bar{n}}{k+p_{F}\bar{n}}, \label{b1}%
\end{equation}
and
\begin{equation}
\left\langle n_{B}\right\rangle |_{n_{F}}=\frac{p_{B}\bar{n}}{k+p_{F}\bar{n}%
}\left(  \frac{k}{2}+n_{F}\right)  . \label{nbm1}%
\end{equation}

This closes the theoretical discussion of the problem.

\section{Results}

In the present section we test our results using the UA5 $p\bar{p}$
forward-backward multiplicity correlation data at $\sqrt{s}=200$ GeV. The
measurement was performed in the pseudorapidity range of $|\eta|<4$ for
various symmetric (around $\eta=0$) forward and backward intervals. In this
case, taking the model with two independent sources, we have $p_{RF}%
=p_{LB}\equiv p$ and $p_{RB}=p_{LF}=1-p$, where probability $p$ is calculated
from Eq. (\ref{p_def}). In the model with only one source we always have
$p_{B}=p_{F}=1/2$. \begin{figure}[h!]
\begin{center}
\includegraphics[scale=1.5]{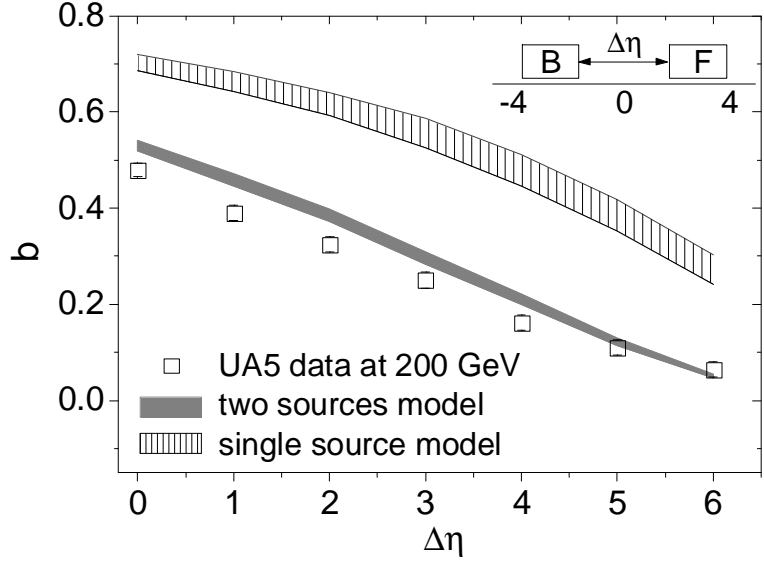}
\end{center}
\caption{The forward-backward multiplicity correlation coefficient $b$ as a
function of the distance $\Delta\eta$ between the backward $B$ and forward $F$
intervals. Data points (squares) measured in $p\bar{p}$ at $\sqrt{s}=200$ GeV
are compared with the results of two models: two independent sources of
particles (grey band) and the model with a single source (dashed band). The
widths of the bands reflect the uncertainty in the value of $k$ from NB fits
to the $p\bar{p}$ multiplicity data.}%
\label{fig_b1}%
\end{figure}\begin{figure}[h!]
\begin{center}
\includegraphics[scale=1.5]{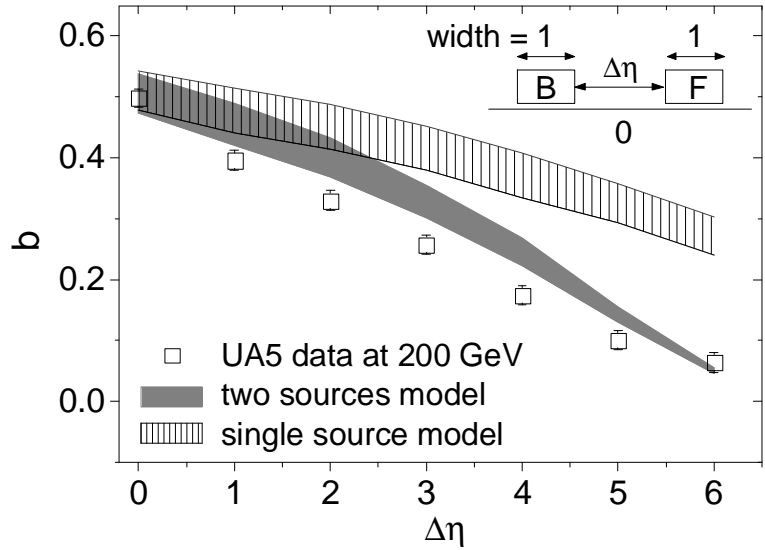}
\end{center}
\caption{The same as in Fig. \ref{fig_b1} but now the width of each interval
is fixed and equals $1$.}%
\label{fig_b2}%
\end{figure}

In Figs. \ref{fig_b1} and \ref{fig_b2} the correlation coefficient $b$ for
various symmetric pseudorapidity intervals is presented. The experimental data
(squares) are taken from Refs. \cite{UA5-fb,NA22-fb}. The grey and dashed
bands represent the results of the model with two independent sources and the
model with a single source, respectively. The widths of the bands reflect the
uncertainty coming from the unknown precise value of $k$ from NB distribution
fits \cite{UA5-200-n-k} to the $p\bar{p}$ multiplicity data. In Fig.
\ref{fig_b1} the forward and backward intervals are chosen as: $B=(-4,-\Delta
\eta/2)$ and $F=(\Delta\eta/2,4)$ with $\Delta\eta=0,1,2,3,4,5,6$. In Fig.
\ref{fig_b2} the forward and backward intervals of constant widths of $1$ are:
$B=(-\Delta\eta/2-1,-\Delta\eta/2)$ and $F=(\Delta\eta/2,\Delta\eta/2+1)$. The
parameters $p$ and $\bar{n}$ are calculated using Eqs. (\ref{p_def}),
(\ref{mn}) and the wounded nucleon fragmentation function shown in Fig.
\ref{fig_wn}. All parameters used in these calculations are listed in Tabs.
\ref{tab1} and \ref{tab2}. In both cases the main source of uncertainties is
the NB parameter $k$, which is not precisely known for all intervals
\cite{UA5-200-n-k}.

As can be observed the model with two independent sources of particles allows
to understand the main features of the data.\ It is worth noticing that the
strong suppression of the correlation coefficient $b$ with increasing
$\Delta\eta$ is fully determined by the suppression of particle production
from a single wounded nucleon to the backward hemisphere. Clearly, the model
in which particles are produced from the single source is incorrect.
\begin{table}[h]
\begin{center}%
\begin{tabular}
[c]{|c|c|c|c|c|}\hline
$\Delta\eta$ & $F$ interval & $\bar{n}$ & $k$ & $p$\\\hline
$0$ & $0.0<\eta<4.0$ & $17.4$ & $3.70\pm0.30$ & $0.70$\\\hline
$1$ & $0.5<\eta<4.0$ & $15.1$ & $3.85\pm0.35$ & $0.73$\\\hline
$2$ & $1.0<\eta<4.0$ & $12.6$ & $3.95\pm0.40$ & $0.76$\\\hline
$3$ & $1.5<\eta<4.0$ & $10.2$ & $4.10\pm0.50$ & $0.80$\\\hline
$4$ & $2.0<\eta<4.0$ & $7.71$ & $4.25\pm0.55$ & $0.84$\\\hline
$5$ & $2.5<\eta<4.0$ & $5.39$ & $4.35\pm0.60$ & $0.89$\\\hline
$6$ & $3.0<\eta<4.0$ & $3.30$ & $4.50\pm0.70$ & $0.94$\\\hline
\end{tabular}
\end{center}
\caption{The parameters used in the calculations of the results presented in
Fig. \ref{fig_b1}.}%
\label{tab1}%
\end{table}\begin{table}[hh]
\begin{center}%
\begin{tabular}
[c]{|c|c|c|c|c|}\hline
$\Delta\eta$ & $F$ interval & $\bar{n}$ & $k$ & $p$\\\hline
$0$ & $0.0<\eta<1.0$ & $4.76$ & $2.30\pm0.30$ & $0.54$\\\hline
$1$ & $0.5<\eta<1.5$ & $4.88$ & $2.70\pm0.40$ & $0.59$\\\hline
$2$ & $1.0<\eta<2.0$ & $4.94$ & $3.05\pm0.45$ & $0.64$\\\hline
$3$ & $1.5<\eta<2.5$ & $4.78$ & $3.40\pm0.50$ & $0.70$\\\hline
$4$ & $2.0<\eta<3.0$ & $4.41$ & $3.80\pm0.60$ & $0.76$\\\hline
$5$ & $2.5<\eta<3.5$ & $3.90$ & $4.10\pm0.60$ & $0.85$\\\hline
$6$ & $3.0<\eta<4.0$ & $3.30$ & $4.50\pm0.70$ & $0.94$\\\hline
\end{tabular}
\end{center}
\caption{The parameters used in the calculations of the results presented in
Fig. \ref{fig_b2}.}%
\label{tab2}%
\end{table}

\begin{figure}[h]
\begin{center}
\includegraphics[scale=1.5]{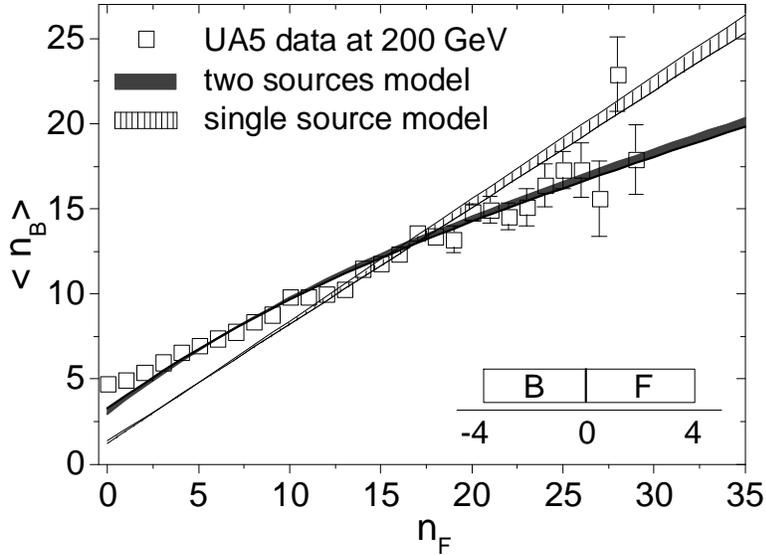}
\end{center}
\caption{The relation between the average number of particles in the backward
interval $\left\langle n_{B}\right\rangle $ at a given number of particles
$n_{F}$ in the forward one. The UA5 $p\bar{p}$ experimental data (squares) at
$\sqrt{s}=200$ GeV are compared with the results of the model with two
independent sources of particles (grey band) and the model with a single
source (dashed band). The widths of the bands reflect the uncertainty in the
value of $k$ from NB fits to the $p\bar{p}$ multiplicity data.}%
\label{fig_line}%
\end{figure}

In Fig. \ref{fig_line} the relation $\left\langle n_{B}\right\rangle |_{n_{F}%
}$ between the average number of particles $\left\langle n_{B}\right\rangle $
in $B$ interval at a given number of particles $n_{F}$ in $F$ interval is
shown. The measurement was performed in two symmetric pseudorapidity intervals
$B=(-4,0)\ $and $F=(0,4)$. Taking Eqs. (\ref{p_def}), (\ref{mn}) into account
we obtain $p=0.7$ and $\bar{n}=17.4$. The main source of uncertainties is the
measured value of $k=3.7\pm0.3$ \cite{UA5-200-n-k}.

The model with two independent sources of particles again correctly describes
the data.\footnote{Except maybe the region of $n_{F}\leqslant3$.} It is
interesting to note that our formalism predicts some deviations from
linearity, which are too small to be noticeable at a given experimental precision.

\section{Conclusions and comments}

Our conclusions can be formulated as follows.

(i) Assuming that in $pp$ collisions \textit{soft} particles are produced from
two independent sources: left- and right-moving wounded nucleons, we have
derived the formulae for the forward-backward multiplicity correlation
coefficient $b$ and the functional relation between the average number of
particles in the backward interval $\left\langle n_{B}\right\rangle $ at a
given number of particles $n_{F}$ in the forward one. This is compared with
the case where only one source contributes to particle spectrum.

(ii) We compared our results with the UA5 $p\bar{p}$ data at $\sqrt{s}=200$
GeV. We conclude that the model with two independent sources of particles
allows to understand the main features of the forward-backward correlation
data. As far as the correlation coefficient is concerned, we observed very
nice qualitative agreement, particularly linear suppression of $b$ with
increasing distance between the forward and backward intervals. This effect is
fully determined by the suppression of the particle production from a wounded
nucleon to the backward hemisphere.

(iii) We also successfully described the functional relation of the average
number of particles in the backward interval $\left\langle n_{B}\right\rangle
$ at a given number $n_{F}$ of particles in the forward one. It is interesting
to note that our formalism predicts some deviations from linearity, which are
too small to be noticeable at a given experimental precision. It would be
interesting to study this effect in the future experiments.

(iv) The model in which the particles are produced from a single source is in
a clear disagreement with the data.

Following comments are in order.

(a) The presented analysis was performed only at $\sqrt{s}=200$ GeV since for
higher energies the wounded nucleon fragmentation functions are unknown.
Studying the correlation data at higher energies should allow to extract these functions.

(b) Assuming $\left\langle n_{B}\right\rangle |_{n_{F}}$ to be in the
parabolic form with the quadratic term $n_{F}^{2}c$ in the range $n_{F}<40$
and $n_{F}<50$ we obtained $c\approx-0.0045$ and $-0.0035$, respectively. It
is interesting to note that qualitatively similar tendency was observed in the
quantum-statistical approach \cite{Fowler}.

\bigskip

\textbf{Acknowledgements}

We would like to thank Andrzej Bia\l as for suggesting this investigation and
useful discussions. This investigation was supported in part by the Polish
Ministry of Science and Higher Education, grant No. N202 034 32/0918.


\begin{thebibliography}{9}
\bibitem {WNM}A. Bialas, M. Bleszynski and W. Czyz, Nucl. Phys. B111 (1976) 461.

\bibitem {PHO-dAu}PHOBOS Collaboration: B.B. Back et al., Phys. Rev. C72
(2005) 031901.

\bibitem {ff-bc}A. Bialas and W. Czyz, Acta Phys. Polon. B36 (2005) 905.

\bibitem {ff-bb}A. Bialas and A. Bzdak, Phys. Rev. C77 (2008) 034908; Phys.
Lett. B649 (2007) 263; Acta Phys. Polon. B38 (2007) 159. For a review, see A.
Bialas, J. Phys. G35 (2008) 044053.

\bibitem {DPM}A. Capella, U. Sukhatme, C-I Tan and J. Tran Thanh Van, Phys.
Rept. 236 (1994) 225.

\bibitem {ff-Ryb}G. Barr, O. Chvala, H.G. Fischer, M. Kreps, M. Makariev, C.
Pattison, A. Rybicki, D. Varga and S. Wenig, Eur. Phys. J. C49 (2007) 919; A.
Rybicki, Acta Phys. Polon. B33 (2002) 1483.

\bibitem {bibzpe}A. Bialas, A. Bzdak and R. Peschanski, Phys. Lett. B665
(2008) 35.

\bibitem {Stod}L. Stodolsky, Phys. Rev. Lett. 28 (1972) 60.

\bibitem {UA5-fb}UA5 Collaboration: R.E. Ansorge et al., Z. Phys. C37 (1988) 191.

\bibitem {FK}K. Fialkowski and A. Kotanski, Phys. Lett. B115 (1982) 425; Phys.
Lett. B107 (1981) 132.

\bibitem {DDD}J. Dias de Deus, Phys. Lett. B100 (1981) 177.

\bibitem {BBP}J. Benecke, A. Bialas and S. Pokorski, Nucl. Phys. B110 (1976)
488, Erratum-ibid. B115 (1976) 547.

\bibitem {Giov}A. Giovannini and R. Ugoccioni, Phys. Rev. D66, (2002) 034001;
Phys. Lett. B558 (2003) 59.

\bibitem {Braun}M.A. Braun, C. Pajares and V.V. Vechernin, Phys. Lett. B493
(2000) 54.

\bibitem {Brogueira}P. Brogueira, J. Dias de Deus and C. Pajares,
arXiv:0901.0997 [hep-ph].

\bibitem {Chou}T.T. Chou and C.N. Yang, Phys. Lett. B135 (1984) 175.

\bibitem {Lim}S.L. Lim, Y.K. Lim, C.H. Oh and K.K. Phua, Z. Phys. C43 (1989)
621; S.L. Lim, C.H. Oh and K.K. Phua, Z. Phys. C54 (1992) 107.

\bibitem {fb-star}A. Bzdak, arXiv:0902.2639 [hep-ph].

\bibitem {NA22-fb}NA22 Collaboration: V.V. Aivazyan et al., Z. Phys. C42
(1989) 533.

\bibitem {UA5-200-n-k}UA5 Collaboration: R.E. Ansorge et al., Z. Phys. C43
(1989) 357.

\bibitem {Fowler}G.N. Fowler, E.M. Friedlander, F.W. Pottarg, R.M. Weiner, J.
Wheeler and G. Wilk, Phys. Rev. D37 (1988) 3127.
\end{thebibliography}
\end{document}